\documentclass[prb,superscriptaddress,twocolumn,floatfix]{revtex4}

\usepackage{graphicx}
\usepackage{dcolumn}
\usepackage{bm}
\usepackage[latin1]{inputenc}
\usepackage{amsmath}
\usepackage{amssymb}

\begin{document}


\title{Microscopic Study of the Superconducting State of the Iron Pnictide RbFe$_{2}$As$_2$}

\author{Z.~Shermadini}
\affiliation{Laboratory for Muon Spin Spectroscopy, Paul Scherrer
Institute, CH-5232 Villigen PSI, Switzerland} \affiliation{Institut
f\"ur Festk\"orperphysik, TU Dresden, D--01069 Dresden, Germany}
\author{J.~Kanter}
 \affiliation{Laboratory for Solid State Physics, ETH Z\"urich, CH-8093 Z\"urich, Switzerland}
\author{C.~Baines}
\affiliation{Laboratory for Muon Spin Spectroscopy, Paul Scherrer
Institute, CH-5232 Villigen PSI, Switzerland}
\author{M.~Bendele}
\affiliation{Laboratory for Muon Spin Spectroscopy, Paul Scherrer
Institute, CH-5232 Villigen PSI, Switzerland}
\affiliation{Physik-Institut der Universit\"at Z\"urich, Winterthurerstrasse 190, CH-8057 Z\"urich, Switzerland}
\author{Z.~Bukowski}
\affiliation{Laboratory for Solid State Physics, ETH Z\"urich, CH-8093
Z\"urich, Switzerland}
\author{R.~Khasanov}
 \affiliation{Laboratory for Muon Spin Spectroscopy, Paul Scherrer
Institute, CH-5232 Villigen PSI, Switzerland}
\author{H.-H.~Klauss}
\affiliation{Institut f\"ur Festk\"orperphysik, TU Dresden, D--01069
Dresden, Germany}
\author{H.~Luetkens}
\affiliation{Laboratory for Muon Spin Spectroscopy, Paul Scherrer
Institute, CH-5232 Villigen PSI, Switzerland}
\author{H.~Maeter}
\affiliation{Institut f\"ur Festk\"orperphysik, TU Dresden, D--01069
Dresden, Germany}
\author{G.~Pascua}
\affiliation{Laboratory for Muon Spin Spectroscopy, Paul Scherrer
Institute, CH-5232 Villigen PSI, Switzerland}
\author{B.~Batlogg}
\affiliation{Laboratory for Solid State Physics, ETH Z\"urich, CH-8093
Z\"urich, Switzerland}
\author{A.~Amato}
\affiliation{Laboratory for Muon Spin Spectroscopy, Paul Scherrer
Institute, CH-5232 Villigen PSI, Switzerland}

\begin{abstract}
A study of the temperature and field dependence of the penetration
depth $\lambda$ of the superconductor RbFe$_{2}$As$_2$
($T_{c}=2.52$~K) was carried out by means of muon-spin rotation
measurements. In addition to the zero temperature value of the
penetration depth $\lambda(0)=267(5)$~nm, a determination of the
upper critical field $B_{c2}(0)=2.6(2)$~T was obtained. The
temperature dependence of the superconducting carrier concentration is discussed within
the framework of a multi-gap scenario. 
Compared to the other ``122'' systems which exhibit much higher Fermi level, a strong reduction of the large gap BCS ratio $2\Delta/k_{\rm B}T_c$ is observed. This
is interpreted as a consequence of the absence of interband processes. Indications of possible pair-breaking effect are also discussed.

\end{abstract}

\maketitle

The iron arsenide AFe$_{2}$As$_{2}$ systems (where A is an alkaline
earth element) crystallize with the tetragonal ThCr$_{2}$Si$_{2}$
type structure (space group $I4/mmm$) \cite{Pfisterer}. The interest
for these compounds arises from the observation of superconductivity
with transition temperatures $T_{c}$ up to 38~K upon alkali metal
substitution for the A element \cite{Rotter, Sasmal, Bukowski}, or
partial transition metal substitution for iron \cite{Sefat}. A huge
number of studies were already devoted to unravel the properties of
their superconducting ground state. However, some studies are
hampered by the fact that to date no clear picture could be drawn
about the bulk character of the superconductivity. For example, in
superconducting systems obtained from the substitution of the A
element (like K for Ba), muon-spin rotation/relaxation ($\mu$SR)
measurements studies clearly indicate the occurrence of phase
separation between magnetic and superconducting phases \cite{Aczel, Park, Khasanov}. On the other hand,
substitution performed on the superconducting plane, as cobalt
substitution for iron, does not reveal any phase separation as
reported also by $\mu$SR \cite{Khasanov2}.

The alkali metal iron arsenide RbFe$_{2}$As$_{2}$ was discovered
some years ago \cite{Czybulka} but was only recently found, by
Bukowski et al. \cite{Bukowski2}, to exhibit type II bulk
superconductivity below $T_c \simeq 2.6$~K. The reported studies
were hindered by a limited temperature range of the equipment and
the full development of the Meissner state could not be recorded.
The estimated value of the upper critical field at zero
temperature, $B_{c2}\simeq2.5$~T, was obtained
from magnetization measurements performed at various field down to
$1.5$~K in the mixed state and by assuming a temperature dependence
provided by the Werthamer--Helfand--Hohenberg theory \cite{Bukowski2}.

Compared to the better known compound
BaFe$_{2}$As$_{2}$, RbFe$_{2}$As$_{2}$ possesses a lower Fermi level
and is characterized by the absence of magnetic instability. 
Furthermore, the electron deficiency in RbFe$_{2}$As$_{2}$ leads also to a
change ({\it i.e.} a decrease) of the number of bands contributing
to the superconducting state, compared for example to
Ba$_{1-x}$K$_{x}$Fe$_{2}$As$_{2}$. Hence, one expects a strong decrease of the contribution of the electron-like bands at the $M$ point of
the Fermi surface. Such a decrease has been observed by angle-resolved photoemission
spectroscopy \cite{Sato2009} in the analog system KFe$_2$As$_2$, which also presents a case
of naturally hole-(over)doped system when compared to the alkaline earth ``122'' iron-based superconductors.

As exemplified by a number of recent studies,
the $\mu$SR technique is very well suited to investigate
the superconducting properties of iron-based
systems (see for example Ref.~\onlinecite{Amato}). In addition,
due to its comparatively low upper critical
field $B_{c2}$ and its reduced $T_c$, the system
RbFe$_2$As$_2$ opens a unique opportunity to fully
study the $B-T$ phase diagram of an iron-arsenide
compound.

In this article, we report on a detailed study of the temperature
and field dependence of the magnetic penetration depth of
RbFe$_2$As$_2$, which is closely related to the superconducting carrier concentration.

Polycrystalline samples of RbFe$_2$As$_2$ were synthesized in
two steps as reported recently \cite{Bukowski2}.
The $\mu$SR measurements were performed at the $\pi$M3 beamline of
the Paul Scherrer Institute (Villigen, Switzerland), using the GPS
instrument (for temperatures down to 1.6~K and field up to 0.6~T) as
well as the LTF instrument (for temperatures down to 0.02~K and
higher fields). Both zero field (ZF) and transverse field (TF)
$\mu$SR measurements were performed. Additional transport studies
were performed on the very same sample at the ETH-Zürich
using an ac Transport Option of a Quantum Design 14T-PPMS.

\begin{figure}[t]
\center{\includegraphics[width=0.8\columnwidth,angle=0,clip]{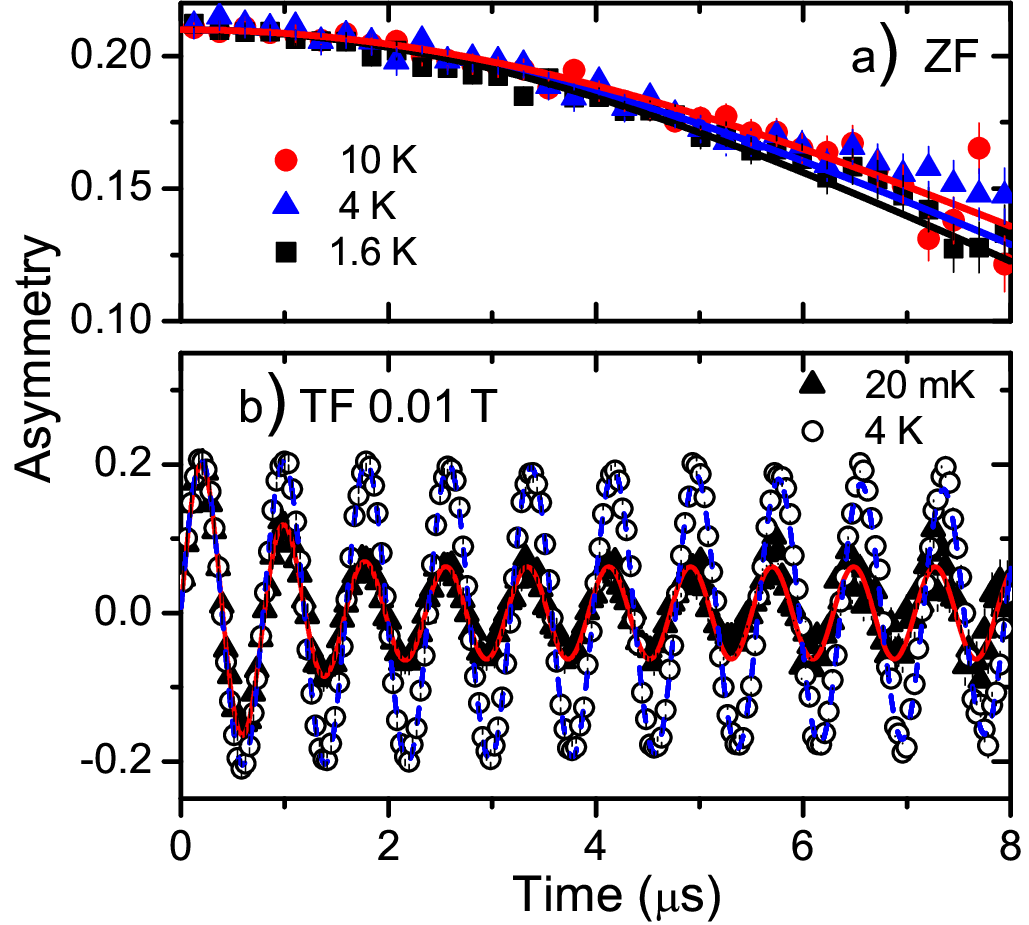}}
    \caption[]{Typical $\mu$SR spectra recorded above and below $T_c$, in: a) zero field; and b) transverse field.} \label{muon_fig}\vspace{-0.3cm}
\end{figure}

To exclude the occurrence of any magnetic contributions of the Fe ions at low temperature, we performed first
ZF mesurements above and below $T_c$. 
As exemplified by the data reported on Fig.~\ref{muon_fig}a, no sign of static magnetism could be detected on the ZF response RbFe$_2$As$_2$. 
The data are well described by a standard Kubo-Toyabe depolarization function \cite{kubo}, reflecting the field distribution at the
muon site created by the nuclear moments.
The marginal increase of the depolarization rate, not related to the superconducting transition, possibly points to a slowing down of the magnetic fluctuations.

Figure~\ref{muon_fig}b exhibits the TF $\mu$SR time spectra
measured in an applied field of 0.01~T, above ($T=4$~K) and below
($T=0.02$~K) the superconducting transition temperature. The strong
muon-spin depolarization at low temperatures reflects the formation
of the flux-line lattice (FLL) in the superconducting state. The
long-lived component detectable at low temperatures is due to a
background contribution from the sample holder. In a
polycrystalline sample the magnetic penetration depth $\lambda$ (and consequently the 
superconducting carrier concentration $n_{\rm s} \propto 1/\lambda^2$) can
be extracted from the Gaussian muon-spin depolarization rate
$\sigma_{\rm s}(T)$ (see also below Eq.~\ref{SigmaSupTheorField}), which reflects the second
moment ($\sigma_{\rm s}^2/\gamma_{\mu}^2$) of the magnetic field
distribution due to the FLL in the mixed state. The TF data were
analyzed using the polarization function:
\begin{eqnarray}
A_{0}P(t)&=& 
A_{\rm s}\exp\left[-\frac{(\sigma_{\rm s}^{2}+\sigma_{\rm n}^{2})t^2}{2}\right]\cos(\gamma_{\mu}B_{\rm int}t+\varphi) \nonumber \\
&&+A_{\rm sh}\exp(-\frac{\sigma_{\rm sh}^{2}t^2}{2})\cos(\gamma_{\mu}B_{\rm sh}t+\varphi)~.
 \label{eq:TF}
 \end{eqnarray}

\begin{figure}[tb]
\center{\includegraphics[width=0.8\columnwidth,angle=0,clip]{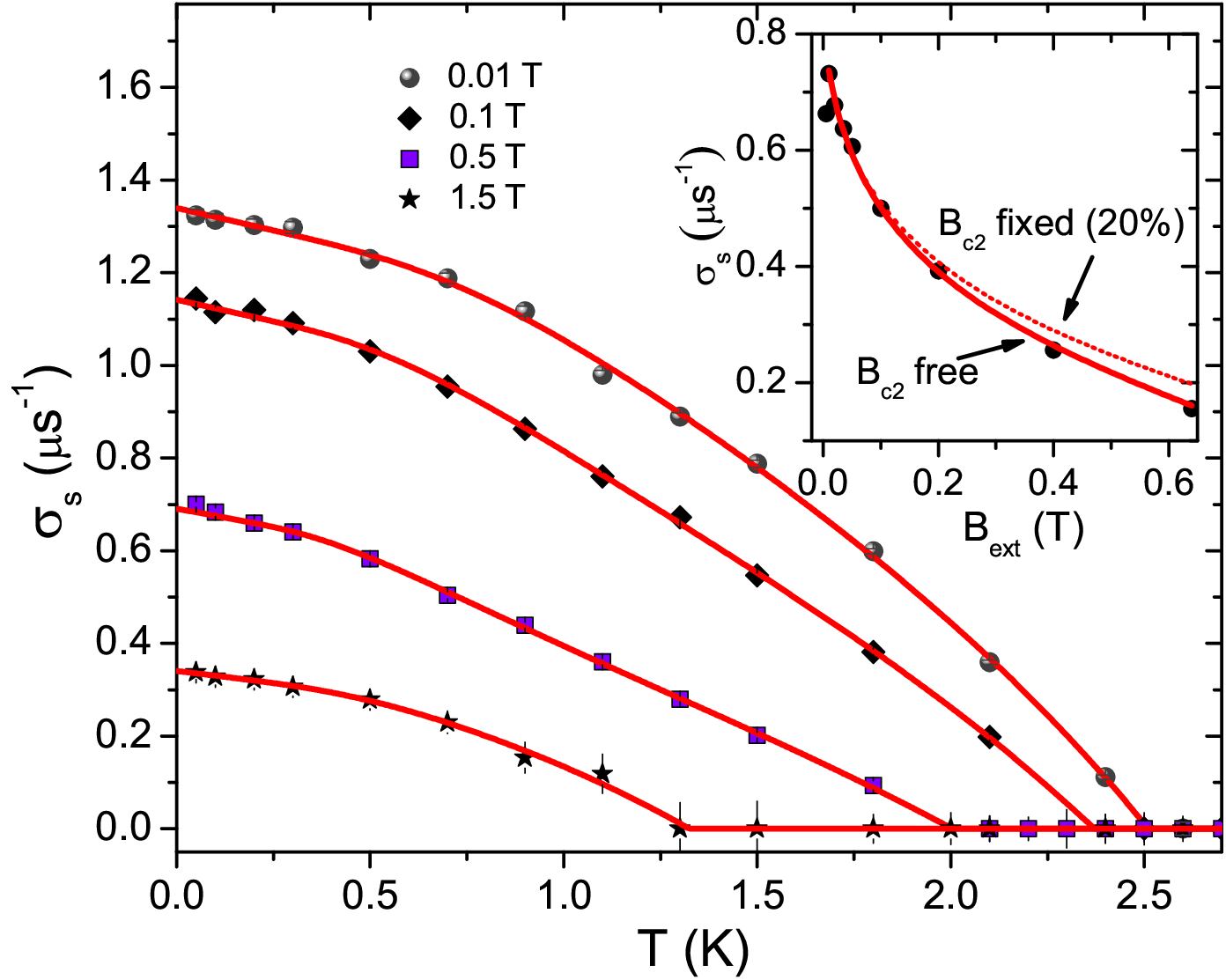}}
    \caption[]{Temperature dependence of the depolarization rate due to the FLL in RbFe$_{2}$As$_{2}$ and obtained in fields of 1.5, 0.5, 0.1 and 0.01~T (lines are guides to the eye). Insert: Field dependence of $\sigma_{\rm s}$ obtained at 1.6~K and analyzed using the Eq.~\ref{SigmaSupTheorField}. The dashed line represents a fit using the $B_{c2}$ value given by the magnetoresistivity data (with an arbitrary criterium of a resistivity increase corresponding to 20\% of its value in the normal state). The solid line corresponds to a similar analysis, but considering $B_{c2}$ as a free parameter.} \label{SigmaSup}\vspace{-0.3cm}
\end{figure}
The first term on the right-hand side of Eq.~\ref{eq:TF} represents
the sample contribution, where $A_{\rm s}$ denotes the initial asymmetry
connected to the sample signal; $\sigma_{\rm s}$ is the  Gaussian
relaxation rate due to the FLL; $\sigma_{\rm n}$ is the contribution to
the field distribution arising from the nuclear moment and which is
found to be temperature independent, in agreement with the ZF results; $B_{\rm int}$ is
the internal magnetic field, sensed by the muons and $\varphi$ is
the initial phase of the muon-spin ensemble. The second term
reflects the muons stopping in the silver sample holder, where
$A_{\rm sh}$ denotes the initial asymmetry connected to the holder
signal; $\sigma_{\rm sh}$ is the  relaxation rate due to the nuclear
moments (which is very close to zero in this case) and $B_{\rm sh}$ is
the magnetic field in the sample holder, which has essentially
the value of the external field. 

On Fig.~\ref{SigmaSup}, we report the temperature dependence of
$\sigma_{\rm s}$ extracted from TF-$\mu$SR measurements in four
different fields. We note first that the perfect fits obtained by assuming a Gaussian field distribution of the FLL point to a rather large
anisotropy of the magnetic penetration depth in our system. This is confirmed by recent $\mu$SR measurements performed on hole- and electron-doped ``122'' systems\cite{Khasanov,Khasanov2}. As expected, $\sigma_{\rm s}$ is zero in the
paramagnetic state and starts to increase below $T_c(B)$ when the
FLL is formed. Upon lowering the temperature,
$\sigma_{\rm s}$ increases gradually reflecting the decrease of the
penetration depth or, alternatively, the increase of the
superconducting density. The overall decrease of $\sigma_{\rm s}$ at very
low temperatures observed upon increasing the applied field is a
direct consequence of the decrease of the width of the internal field
distribution when increasing the field towards $B_{c2}$. In order to
quantify such an effect, one can make use of
the numerical Ginzburg-Landau model, developed by Brandt
\cite{Brandt}. This model allows one to calculate the superconducting carrier concentration with good approximation within the local (London)
approximation ($\lambda \gg \xi$, $\xi$ is the coherence length).
This model predicts the  magnetic field dependence of the second
moment of the magnetic field distribution or, alternatively, of the
$\mu$SR depolarization rate, which can be expressed as:

\begin{equation}
\begin{array}{rl}
\sigma_{\rm s}\;[\mu\text{\rm s}^{-1}]=&4.83 \times 10^{4}(1-B/B_{c2})\times\\
&\times[1+1.21(1-\sqrt{B/B_{c2}})^{3}]\lambda^{-2}\;[\text{nm}]\label{SigmaSupTheorField}
\end{array}
\end{equation}

\begin{figure}[tb]
\center{\includegraphics[width=0.8\columnwidth,angle=0,clip]{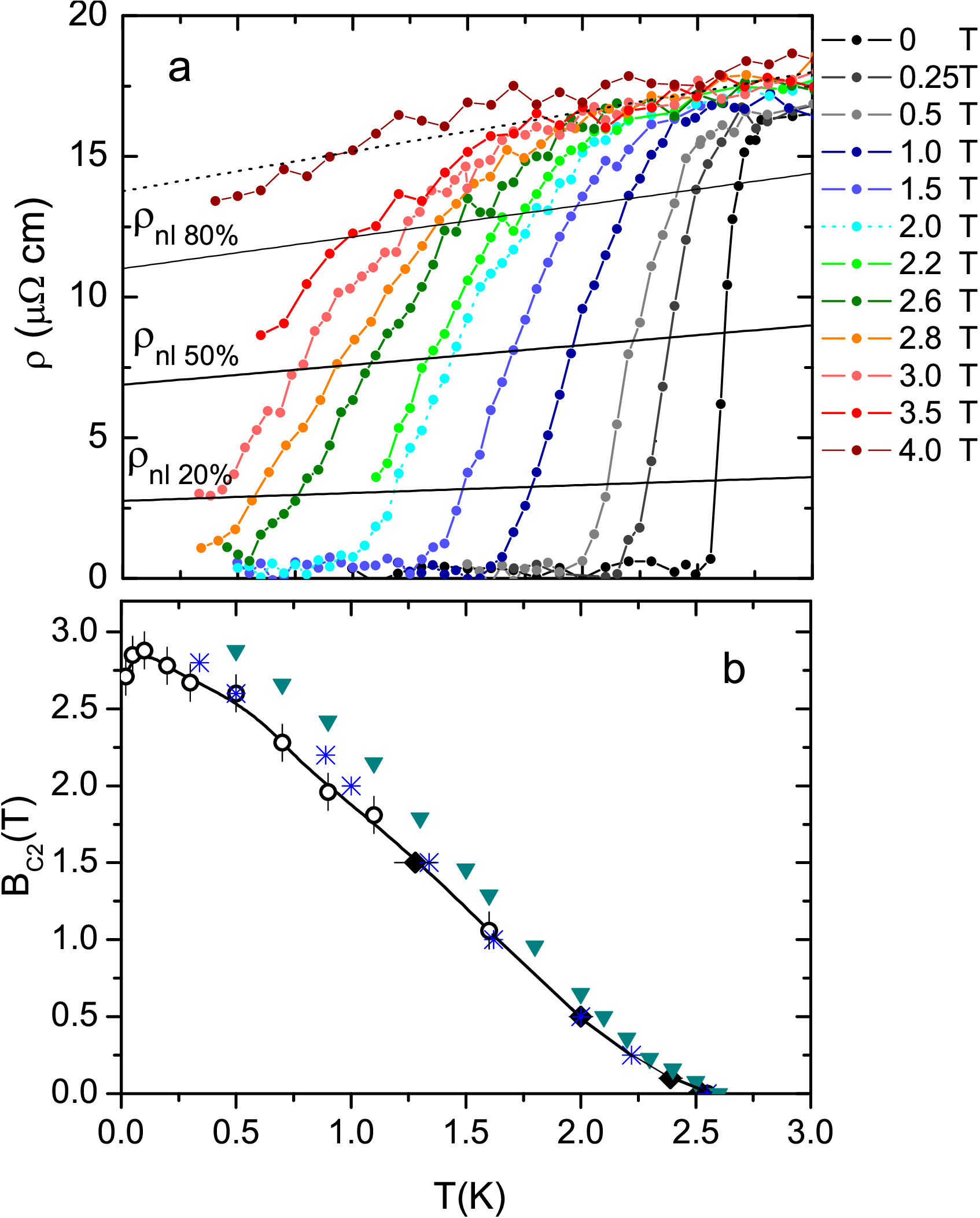}}
    \caption[]{a) Field dependence of the electrical resistivity.
    b) Upper critical field for RbFe$_2$As$_2$. The open circles are obtained by analyzing the field dependence of $\sigma_{\rm s}$ using Eq.~\ref{SigmaSupTheorField}, as explained in the text.
    The diamonds are the value obtained by analyzing the temperature dependence of $\sigma_{\rm s}$. The blue stars correspond to the complete disappearence of the resistivity in field.
    For comparison, we report the values given by the magnetoresistivity data with the same criterium used in the insert of Fig.~\ref{SigmaSup} and corresponding to the values extracted along the line $\rho_\text{nl 20\%}$ of the upper panel.
    The line is a guide to the eye.} \label{RhoField}\vspace{-0.5cm}
\end{figure}

\begin{figure}[htbp]
\center{\includegraphics[width=0.8\columnwidth,angle=0,clip]{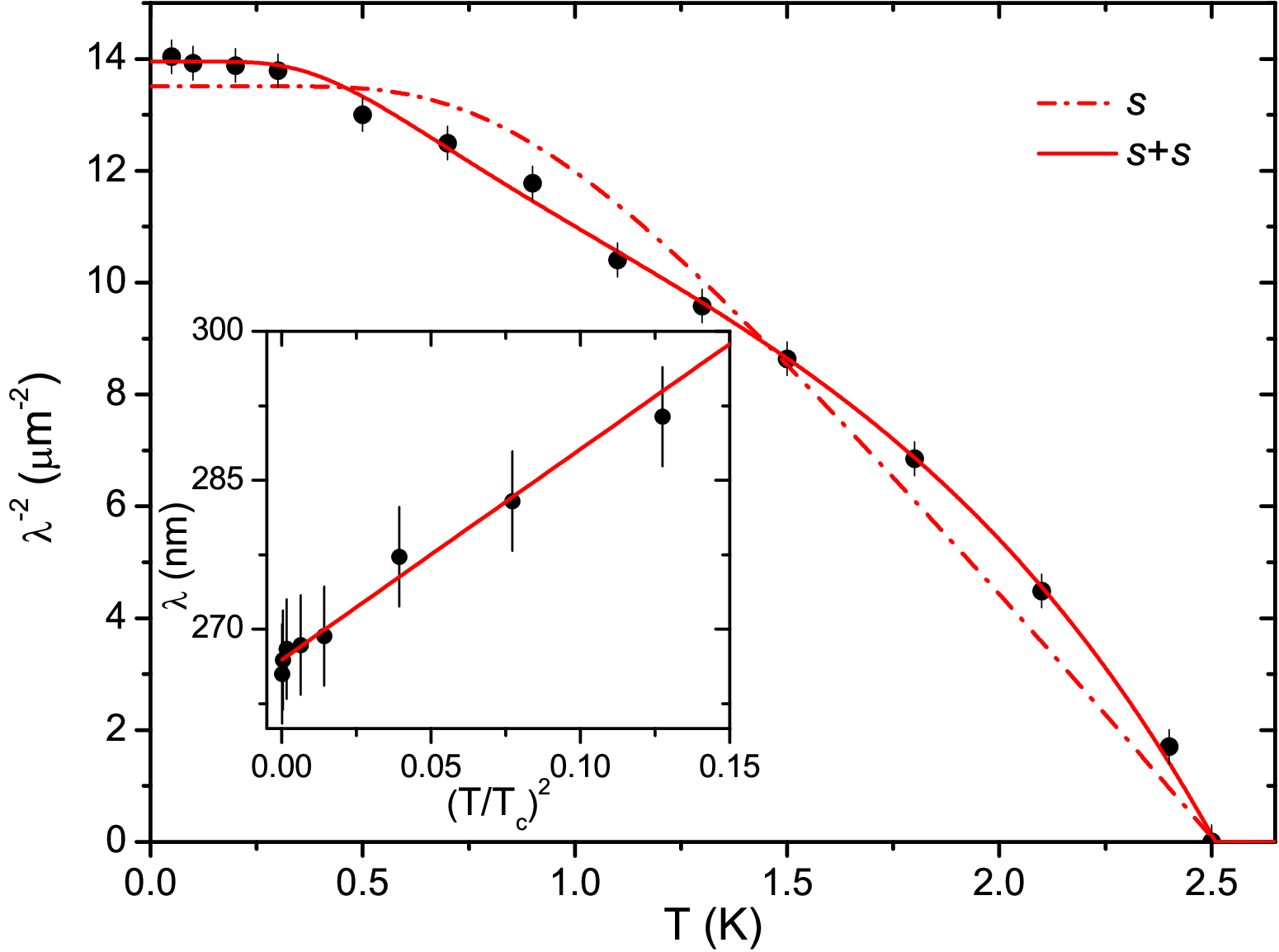}}
    \caption[]{Magnetic penetration depth as a function of temperature. Above 0.5~K, the values
    obtained with Eq.~\ref{SigmaSupTheorField} coincides with the values measured in a field
    of 0.01~T, and only these latter are plotted for this temperature range.
    The red dashed line corresponds to a BCS $s$-wave gap symmetry, whereas the solid one to represents a fit using a two-gap $s+s$ model. The insert exhibits the penetration depth as a function of $(T/T_{c})^{2}$.} \label{PenetrationDepth}\vspace{-0.3cm}
\end{figure}

The insert of Fig.~\ref{SigmaSup} exhibits the evolution of $\sigma_{\rm s}$ at
1.6~K as a function of the external applied magnetic field. For each
data point, the sample was field-cooled from above $T_{c}$ down to
1.6~K and the recorded $\mu$SR spectra were analyzed with
Eq.~\ref{eq:TF}. The field dependence of $\sigma_{\rm s}$ was 
analyzed with Eq.~\ref{SigmaSupTheorField} using the values of the
upper critical field $B_{c2}$ given either by an arbitrary criterium
based on magnetoresistivity measurements (see Fig.~\ref{RhoField}a)
or by leaving the parameter $B_{c2}$ free. The latter option
provides excellent fits for all temperatures and the corresponding
fitted values of the superconducting carrier concentration and of the upper critical fields 
are reported on Fig.~\ref{PenetrationDepth} and \ref{RhoField}b. An additional
point provided by this first investigation, is that the penetration depth can
be assumed to be field independent. This rules out the possibility that RbFe$_2$As$_2$ is a nodal superconductor,
since a field should have induced excitations at the gap
nodes due to nonlocal and nonlinear effects, thus reducing
the superconducting carrier concentration $n_{\rm s}$ and therefore affecting $\lambda$ (see for example Ref.~\onlinecite{Amin2000}).  

By looking at the temperature dependence of $\lambda^{-2}$ obtained using Eq.~\ref{SigmaSupTheorField} with the values of the parameter $B_{c2}(T)$ presented on Fig.~\ref{RhoField}b, 
the zero-temperature value of the penetration depth
$\lambda(0)=267(5)$~nm can be deduced. The obtained temperature dependence of $\lambda^{-2}$ was analyzed, in a first step, within the framework of a BCS single $s$-wave symmetry superconducting gap $\Delta$ (see Fig.~\ref{PenetrationDepth}), using the form\cite{Tinkham}:
\begin{equation}
\frac{\lambda^{-2}(T)}{\lambda^{-2}(0)}=1-\frac{2}{k_BT}\int_\Delta^{\infty} f(\epsilon ,T)\left[ 1-f(\epsilon ,T)\right] d\epsilon~,\label{Lambda2}
\end{equation}
where $f(\epsilon ,T) = ( 1+\exp [\sqrt{\epsilon^2+\Delta(T)^2}/k_BT])^{-1}$,
and with a standard BCS temperature dependence for the gap function.  
As evidenced on Fig.~\ref{PenetrationDepth}, this analysis is not satisfactory. We note also that a $d$-wave symmetry model does not fit the data, confirming at posteriori the discussion of a field independent penetration depth. These results are actually not unexpected, as there are growing evidences that several disconnected Fermi-surface sheets contribute to the superconductivity, as revealed by angle-resolved photoemission
spectroscopy \cite{Sato2009}, resulting into two distinct values of superconducting gaps. Hence, in a second step, the experimental $\lambda^{-2}(T)$ data were analyzed by assuming two independent contributions with different values $\Delta_i$ of $s$-wave gaps \cite{Niedermayer, Carrington, Khasanov}.
On Fig.~\ref{PenetrationDepth} the solid line shows a $s+s$ multi-gap function which fits to the experimental data
rather well. The parameters extracted from the fit are $\Delta_1(0) = 0.15(2)$~meV for the small gap value (contributing $\omega=36\%$ to the total amount of $n_{\rm s}$) and $\Delta_2 = 0.49(4)$~meV for the larger one. However, note that according to Eq.~\ref{Lambda2}, $\lambda^{-2}$ is insensitive to the phase of the superconducting gap(s). By considering the intrinsic hole-doping in RbFe$_2$As$_2$ compared to the optimally doped ``122'' iron-based system, it is natural to consider that the gaps values are connected respectively to the outer ($\beta$) and inner ($\alpha$) hole-like bands at the $\Gamma$ point of the Fermi surface. In this frame, RbFe$_2$As$_2$ can be considered as hole-overdoped with  electron-like $\gamma$ and $\delta$ bands at the $M$ point, which shift to the unoccupied side. Note that in optimally doped ``122'' systems, one observes the occurrence of $\epsilon$ hole bands (so-called ``blades'') around the $M$ point, which also slightly contribute to the superconducting carrier concentration.

An additional support for a two-gap superconducting state could
be provided by 
the observed positive
curvature of the $B_{c2}(T)$ near $T_c$, in sharp contrast to
the usual $B_{c2}$ BCS temperature dependence (see Fig.~\ref{RhoField}b). 
Note first that the values of $B_{c2}$ extracted
from the fit with Eq.~\ref{SigmaSupTheorField} are in perfect
agreement with: i) the values corresponding to the complete
suppression of the electrical resistivity in field; and ii) to the
values obtained by analyzing the temperature dependence of
$\sigma_{\rm s}$ in different magnetic fields (see Fig.~\ref{SigmaSup}).
An additional indication that bulk superconductivity occurs when the
electrical resistivity completely vanishes is provided by specific
heat measurements \cite{jakob} performed in zero-applied field for
which the observed $T_c$ corresponds to $2.50(1)$~K.

Similar positive curvature of the $B_{c2}(T)$ near $T_c$ were observed in MgB$_2$ \cite{deLima2001,Sologubenko2002} and in the borocarbides \cite{shulga1998}, where it was explained within a two-gap model. However, one should keep in mind that alternative explanations for the observed positive curvature in $B_{c2}(T)$ are possible and that complementary measurements, as here our $\lambda^{-2}(T)$ data, are necessary to draw conclusions.

If on one hand, the two-gap model scenario appears to best fit the temperature dependence of the penetration depth, on the other hand one could argue that it does not appear fully consistent with the observation that the field evolution of the field distribution follows Eq.~\ref{SigmaSupTheorField}. Hence for a two-gap model, one expects a deviation from the simple field dependence reflecting the occurrence of distinct lengths scales $\xi_i$ for both gaps (associated to the coherence length, for a clean single gap system). Such behavior is for example clearly observed on the archetypical two-gap superconductor MgB$_2$ \cite{serventi}. The experimental observation that Eq.~\ref{SigmaSupTheorField} reproduces our data indicates a small difference between the $\xi_i$ parameters for both bands. This is also inline with the very good agreement between the extracted values of $B_{c2}$ with Eq.~\ref{SigmaSupTheorField} and the observed values by resistivity. In this frame, we also mention that ARPES measurements\cite{Evtushinsky2009} on members of the ``122'' family indicate that the Fermi velocity of the inner $\Gamma$-barrel band ($\alpha$ band) is substantially higher that the one for the outer $\Gamma$-barrel band ($\beta$ band), which therefore weakens the difference of the gap values on the $\xi_i$ parameters (as $\xi \propto \langle v_{\rm F}\rangle/\Delta$). Finally, we note that the observed depolarization rate in RbFe$_2$As$_2$ is about 40 times weaker than the one reported for MgB$_2$, hampering therefore the determination of possible distinct $\xi_i$ length scales.

For completeness, we discuss now the slight deviation observed at very low temperatures from the $s+s$ fit and the $\lambda^{-2}(T)$ data. 
Recently, it was shown that the observation of universal scalings in the whole iron-pnictides superconductors, for the specific
heat jump ($\Delta C \propto T_{c}^{3}$) and the slope of upper critical field at $T_{c}$ ($dB_{c2}/dT \propto T_c$) could be interpreted as signatures for strong pair-breaking effects\cite{Kogan}, as for example magnetic scattering. In the same frame it was deduced\cite{Gordon2} that such an effect should lead to a very low temperature dependence of the penetration depth deviating from an usual exponential behavior and transforming into a quadratic one, i.e. $\lambda \propto T^2$, which is indeed reported in a number of studies (see for example Ref.~\onlinecite{Martin, Kim}).
On the insert of Fig.~\ref{PenetrationDepth} we report the extracted penetration depth as a function of $(T/T_{c})^{2}$. 
The good scaling is inline with the presence of magnetic scattering in RbFe$_2$As$_2$, as previously reported for hole- or electron-doped ``122'' systems \cite{Gordon2}.

To conclude, $\mu$SR measurements were performed on a
RbFe$_{2}$As$_{2}$ polycrystalline sample. From the temperature and
field dependence of the superconducting response of the $\mu$SR
signal, the values of the upper critical field and of the magnetic
penetration depth could be extracted.  The zero temperature values
of $B_{c2}(0)$ and $\lambda(0)$ were estimated to be $2.6(2)$~T
and $267(5)$~nm, respectively. The temperature dependence of the
penetration depth, and similarly of the superconducting carrier concentration, are
reproduced assuming a multi-gap model, with possibly pair-breaking effects at low temperatures.
The multi-gap scenario is supported by the observation of a clear
positive curvature on the temperature dependence of the upper
critical field. We attribute these gaps to the hole-like bands around the $\Gamma$ point of the Fermi surface, and possibly also to the hole-bands blades around the $M$ point. Assuming that the $\gamma$ and $\delta$ electron-like bands around the $M$ point are in the unoccupied side, one would expect an absence of nesting conditions in RbFe$_2$As$_2$. The consequence would be an absence of magnetic order, as confirmed by our ZF data, and a strong decrease of the interband processes between the $\alpha$ and $\gamma$($\delta$) bands. In this frame, it is remarkable to see that the ratio between the gaps values is decreased by a factor more than 2 compared to optimally doped ``122'' systems. Similarly, we note that the BCS ratio $2\Delta/k_{\rm B}T_c$ for the small gap that we assign to the $\beta$ band is almost identical to the values observed for optimally doped Ba$_{1-x}$K$_x$Fe$_2$As$_2$, i.e. $2\Delta_1/k_{\rm B}T_c \simeq 1.4$. On the other side, for the large gap of the $\alpha$ band, this ratio is strongly reduced \cite{Khasanov,Evtushinsky2009}, confirming therefore the possible role played by interband processes in optimally hole-doped iron-based ``122'' superconductors.
 
Part of this work was performed at the Swiss Muon Source (S$\mu$S),
Paul Scherrer Institute (PSI, Switzerland). The work of M.B. was supported
by the Swiss National Science Foundation. The
work at the IFW Dresden has been supported by the
DFG through FOR 538.

\begin{thebibliography}{29}
\expandafter\ifx\csname natexlab\endcsname\relax\def\natexlab#1{#1}\fi
\expandafter\ifx\csname bibnamefont\endcsname\relax
  \def\bibnamefont#1{#1}\fi
\expandafter\ifx\csname bibfnamefont\endcsname\relax
  \def\bibfnamefont#1{#1}\fi
\expandafter\ifx\csname citenamefont\endcsname\relax
  \def\citenamefont#1{#1}\fi
\expandafter\ifx\csname url\endcsname\relax
  \def\url#1{\texttt{#1}}\fi
\expandafter\ifx\csname urlprefix\endcsname\relax\def\urlprefix{URL }\fi
\providecommand{\bibinfo}[2]{#2}
\providecommand{\eprint}[2][]{\url{#2}}

\bibitem[{\citenamefont{Pfisterer and Nagorsen}(1980)}]{Pfisterer}
\bibinfo{author}{\bibfnamefont{M.}~\bibnamefont{Pfisterer}} \bibnamefont{and}
  \bibinfo{author}{\bibfnamefont{G.}~\bibnamefont{Nagorsen}},
  \bibinfo{journal}{Z. Naturforsch.} \textbf{\bibinfo{volume}{35}}
  (\bibinfo{year}{1980}).

\bibitem[{\citenamefont{Rotter et~al.}(2008)}]{Rotter}
\bibinfo{author}{\bibfnamefont{M.}~\bibnamefont{Rotter}} \bibnamefont{et~al.},
  \bibinfo{journal}{Phys.\ Rev.\ Lett.} \textbf{\bibinfo{volume}{101}},
  \bibinfo{pages}{107006} (\bibinfo{year}{2008}).

\bibitem[{\citenamefont{Sasmal et~al.}(2008)}]{Sasmal}
\bibinfo{author}{\bibfnamefont{K.}~\bibnamefont{Sasmal}} \bibnamefont{et~al.},
  \bibinfo{journal}{Phys.\ Rev.\ Lett.} \textbf{\bibinfo{volume}{101}},
  \bibinfo{pages}{107007} (\bibinfo{year}{2008}).

\bibitem[{\citenamefont{Bukowski et~al.}(2009{\natexlab{a}})}]{Bukowski}
\bibinfo{author}{\bibfnamefont{Z.}~\bibnamefont{Bukowski}}
  \bibnamefont{et~al.}, \bibinfo{journal}{Phys.\ Rev.\ B}
  \textbf{\bibinfo{volume}{79}}, \bibinfo{pages}{104521}
  (\bibinfo{year}{2009}{\natexlab{a}}).

\bibitem[{\citenamefont{Sefat et~al.}(2008)}]{Sefat}
\bibinfo{author}{\bibfnamefont{A.~S.} \bibnamefont{Sefat}}
  \bibnamefont{et~al.}, \bibinfo{journal}{Phys.\ Rev.\ Lett.}
  \textbf{\bibinfo{volume}{101}}, \bibinfo{pages}{117004}
  (\bibinfo{year}{2008}).

\bibitem[{\citenamefont{Aczel et~al.}(2008)}]{Aczel}
\bibinfo{author}{\bibfnamefont{A.~A.} \bibnamefont{Aczel}}
  \bibnamefont{et~al.}, \bibinfo{journal}{Phys. Rev. B}
  \textbf{\bibinfo{volume}{78}}, \bibinfo{pages}{214503}
  (\bibinfo{year}{2008}).

\bibitem[{\citenamefont{Park et~al.}(2009)}]{Park}
\bibinfo{author}{\bibfnamefont{J.~T.} \bibnamefont{Park}} \bibnamefont{et~al.},
  \bibinfo{journal}{Phys.\ Rev.\ Lett.} \textbf{\bibinfo{volume}{102}},
  \bibinfo{pages}{117006} (\bibinfo{year}{2009}).

\bibitem[{\citenamefont{Khasanov et~al.}(2009{\natexlab{a}})}]{Khasanov}
\bibinfo{author}{\bibfnamefont{R.}~\bibnamefont{Khasanov}}
  \bibnamefont{et~al.}, \bibinfo{journal}{Phys. Rev. Lett.}
  \textbf{\bibinfo{volume}{102}}, \bibinfo{pages}{187005}
  (\bibinfo{year}{2009}{\natexlab{a}}).

\bibitem[{\citenamefont{Khasanov et~al.}(2009{\natexlab{b}})}]{Khasanov2}
\bibinfo{author}{\bibfnamefont{R.}~\bibnamefont{Khasanov}}
  \bibnamefont{et~al.}, \bibinfo{journal}{Phys. Rev. Lett.}
  \textbf{\bibinfo{volume}{103}}, \bibinfo{pages}{067010}
  (\bibinfo{year}{2009}{\natexlab{b}}).

\bibitem[{\citenamefont{Czybulka et~al.}(1992)}]{Czybulka}
\bibinfo{author}{\bibfnamefont{A.}~\bibnamefont{Czybulka}}
  \bibnamefont{et~al.}, \bibinfo{journal}{Z. Anorg. Allg. Chem.}
  \textbf{\bibinfo{volume}{122}}, \bibinfo{pages}{609} (\bibinfo{year}{1992}).

\bibitem[{\citenamefont{Bukowski et~al.}(2009{\natexlab{b}})}]{Bukowski2}
\bibinfo{author}{\bibfnamefont{Z.}~\bibnamefont{Bukowski}}
  \bibnamefont{et~al.}, \bibinfo{journal}{arXiv:0909.2740v1}
  (\bibinfo{year}{2009}{\natexlab{b}}).

\bibitem[{\citenamefont{Sato et~al.}(2009)}]{Sato2009}
\bibinfo{author}{\bibfnamefont{T.}~\bibnamefont{Sato}} \bibnamefont{et~al.},
  \bibinfo{journal}{Phys.\ Rev.\ Lett.} \textbf{\bibinfo{volume}{103}},
  \bibinfo{pages}{047002} (\bibinfo{year}{2009}).

\bibitem[{\citenamefont{Amato et~al.}(2009)}]{Amato}
\bibinfo{author}{\bibfnamefont{A.}~\bibnamefont{Amato}} \bibnamefont{et~al.},
  \bibinfo{journal}{Physica.\ C} \textbf{\bibinfo{volume}{469}},
  \bibinfo{pages}{606} (\bibinfo{year}{2009}).

\bibitem[{\citenamefont{Kubo and Toyabe}(1967)}]{kubo}
\bibinfo{author}{\bibfnamefont{R.}~\bibnamefont{Kubo}} \bibnamefont{and}
  \bibinfo{author}{\bibfnamefont{T.}~\bibnamefont{Toyabe}},
  \emph{\bibinfo{title}{Magnetic Resonance and Relaxation}}
  (\bibinfo{publisher}{North Holland, Amsterdam}, \bibinfo{year}{1967}).

\bibitem[{\citenamefont{Brandt}(2003)}]{Brandt}
\bibinfo{author}{\bibfnamefont{E.~H.} \bibnamefont{Brandt}},
  \bibinfo{journal}{Phys.\ Rev.\ B} \textbf{\bibinfo{volume}{68}},
  \bibinfo{pages}{054506} (\bibinfo{year}{2003}).

\bibitem[{\citenamefont{Amin et~al.}(2000)}]{Amin2000}
\bibinfo{author}{\bibfnamefont{M.}~\bibnamefont{Amin}} \bibnamefont{et~al.},
  \bibinfo{journal}{Phys.\ Rev.\ Lett.} \textbf{\bibinfo{volume}{84}},
  \bibinfo{pages}{5864} (\bibinfo{year}{2000}).

\bibitem[{\citenamefont{Tinkham}(1996)}]{Tinkham}
\bibinfo{author}{\bibfnamefont{M.}~\bibnamefont{Tinkham}},
  \emph{\bibinfo{title}{Introduction to Superconductivity}}
  (\bibinfo{publisher}{McGraw-Hill Inc.}, \bibinfo{year}{1996}).

\bibitem[{\citenamefont{Niedermayer et~al.}(2002)}]{Niedermayer}
\bibinfo{author}{\bibfnamefont{C.}~\bibnamefont{Niedermayer}}
  \bibnamefont{et~al.}, \bibinfo{journal}{Phys. Rev. B}
  \textbf{\bibinfo{volume}{65}}, \bibinfo{pages}{094512}
  (\bibinfo{year}{2002}).

\bibitem[{\citenamefont{Carrington and Manzano}(2003)}]{Carrington}
\bibinfo{author}{\bibfnamefont{A.}~\bibnamefont{Carrington}} \bibnamefont{and}
  \bibinfo{author}{\bibfnamefont{F.}~\bibnamefont{Manzano}},
  \bibinfo{journal}{Physica C} \textbf{\bibinfo{volume}{385}},
  \bibinfo{pages}{205} (\bibinfo{year}{2003}).

\bibitem[{\citenamefont{Kanter}()}]{jakob}
\bibinfo{author}{\bibfnamefont{J.}~\bibnamefont{Kanter}},
  \bibinfo{note}{private communication}.

\bibitem[{\citenamefont{de~Lima et~al.}(2001)}]{deLima2001}
\bibinfo{author}{\bibfnamefont{O.}~\bibnamefont{de~Lima}} \bibnamefont{et~al.},
  \bibinfo{journal}{Phys.\ Rev.\ Lett.} \textbf{\bibinfo{volume}{86}},
  \bibinfo{pages}{5974} (\bibinfo{year}{2001}).

\bibitem[{\citenamefont{Sologubenko et~al.}(2002)}]{Sologubenko2002}
\bibinfo{author}{\bibfnamefont{A.}~\bibnamefont{Sologubenko}}
  \bibnamefont{et~al.}, \bibinfo{journal}{Phys.\ Rev.\ B}
  \textbf{\bibinfo{volume}{65}}, \bibinfo{pages}{180505}
  (\bibinfo{year}{2002}).

\bibitem[{\citenamefont{Shulga et~al.}(1998)}]{shulga1998}
\bibinfo{author}{\bibfnamefont{S.}~\bibnamefont{Shulga}} \bibnamefont{et~al.},
  \bibinfo{journal}{Phys.\ Rev.\ Lett.} \textbf{\bibinfo{volume}{80}},
  \bibinfo{pages}{1730} (\bibinfo{year}{1998}).

\bibitem[{\citenamefont{Serventi et~al.}(2004)}]{serventi}
\bibinfo{author}{\bibfnamefont{S.}~\bibnamefont{Serventi}}
  \bibnamefont{et~al.}, \bibinfo{journal}{Phys.\ Rev.\ Lett.}
  \textbf{\bibinfo{volume}{93}}, \bibinfo{pages}{217003}
  (\bibinfo{year}{2004}).

\bibitem[{\citenamefont{Evtushinsky et~al.}(2010)}]{Evtushinsky2009}
\bibinfo{author}{\bibfnamefont{D.~V.} \bibnamefont{Evtushinsky}}
  \bibnamefont{et~al.}, \bibinfo{journal}{New J. Phys.}
  \textbf{\bibinfo{volume}{11}}, \bibinfo{pages}{180409(R)}
  (\bibinfo{year}{2010}).

\bibitem[{\citenamefont{Kogan}(2009)}]{Kogan}
\bibinfo{author}{\bibfnamefont{V.~G.} \bibnamefont{Kogan}},
  \bibinfo{journal}{Phys. Rev. B} \textbf{\bibinfo{volume}{80}},
  \bibinfo{pages}{214532} (\bibinfo{year}{2009}).

\bibitem[{\citenamefont{Gordon et~al.}(2010)}]{Gordon2}
\bibinfo{author}{\bibfnamefont{R.~T.} \bibnamefont{Gordon}}
  \bibnamefont{et~al.}, \bibinfo{journal}{arXiv:0912.5346}
  (\bibinfo{year}{2010}).

\bibitem[{\citenamefont{Martin et~al.}(2010)}]{Martin}
\bibinfo{author}{\bibfnamefont{C.}~\bibnamefont{Martin}} \bibnamefont{et~al.},
  \bibinfo{journal}{Phys.\ Rev.\ Lett.} \textbf{\bibinfo{volume}{102}},
  \bibinfo{pages}{247002} (\bibinfo{year}{2010}).

\bibitem[{\citenamefont{Kim et~al.}(2010)}]{Kim}
\bibinfo{author}{\bibfnamefont{H.}~\bibnamefont{Kim}} \bibnamefont{et~al.},
  \bibinfo{journal}{arXiv:1001.2042v1}  (\bibinfo{year}{2010}).

\end{thebibliography}

\end{document}